\documentstyle[eqsecnum,pre,preprint,aps]{revtex}
\tightenlines

\newcommand{\be}{\begin{equation}}

\newcommand{\ee}{\end{equation}}
\newcommand{\bs}{\begin{mathletters}} 
\newcommand{\es}{\end{mathletters}} 

\newcommand{\baa}{\begin{eqnarray}}
\newcommand{\eaa}{\end{eqnarray}}

\newcommand{\ba}{\bs\begin{eqnarray}}
\newcommand{\ea}{\end{eqnarray}\es}

\newcommand{\bt}[1]{\bs\label{#1}\begin{eqnarray}}
\newcommand{\et}{\end{eqnarray}\es}

\newcommand{\figlab}[2]{\begin{figure}\caption{#2}\label{#1}\end{figure}}

\newcommand{\der}[2]{\frac{\partial #1}{\partial #2}} 
\newcommand{\eq}[1]{Eq.~(\ref{#1})}

\newcommand{\paper}[6]{#1 , #2 #3 {\bf #4} , #5 (19#6)}

\newcommand{\refsec}[1]{Sec.~\ref{#1}}

\newcommand{\reffig}[1]{Fig.~\ref{#1}}

\newcommand{\ddd}{{\cal D}}

\newcommand{\vev}[1]{\left\langle\ #1\ \right\rangle}
\newcommand{\half}{\frac{1}{2}}
\newcommand{\psib}{\bar{\psi}}
\newcommand{\derf}[2]{\frac{\delta #1}{\delta #2}}
\newcommand{\dd}[2]{\frac{d #1}{d #2}}
\newcommand{\zb}{\bar{z}}

\newlength{\www}

\newcommand{\smooth}{C^\infty_p(\left[0,\beta\right])}

\begin{document}
\draft
\preprint{ULDF-TH-1/4/98}

\title{Simulation of Supersymmetric Models\\
with a Local Nicolai Map}

\author{
Matteo Beccaria$^{(1)}$, Giuseppe Curci$^{(2)}$, Erika D'Ambrosio$^{(2)}$}

\address{${}^1$ 
Dipartimento di Fisica dell'Universit\`a di Lecce, I-73100, Italy,\\
Istituto Nazionale di Fisica Nucleare, Sezione di Lecce}

\address{${}^2$ 
Dipartimento di Fisica dell'Universit\'a di Pisa, I-56100, Italy, \\
Istituto Nazionale di Fisica Nucleare, Sezione di Pisa}

\maketitle

\begin{abstract}
We study the numerical simulation of supersymmetric models having a local
Nicolai map. The mapping can be regarded as a stochastic equation and its
numerical integration provides an algorithm for the simulation of the
original model.
In this paper, the method is discussed in details and applied to 
examples in $0+1$ and $1+1$ dimensions.
\end{abstract}


\section{Introduction}

A deep property of supersymmetric field theories is the existence
of the Nicolai map~\cite{Nicolai1} that is a non linear transformation of
the bosonic field $\phi$ such that: (i) the transformed action describes 
a gaussian field $\xi$ with unit covariance, (ii) 
the Jacobian determinant of
the transformation exactly cancels the fermion determinant.

In some cases, the map is local: the new field $\xi$ is
expressed by a polynomial in $\phi$ and its derivatives. 
When this happens, it is convenient to regard $\xi$ as a random field
and the transformation $\xi = \xi(\phi)$ 
as a stochastic equation~\cite{Kampen}.

A $0+1$ dimensional example is 
Supersymmetric Quantum Mechanics~\cite{Witten1} which we shall
adopt as a toy model in later discussions. 
In $1+1$ dimensions, a class of models with local Nicolai map is 
that of $N=2$ Wess-Zumino (WZ) models. Other examples are 
$1+3$ dimensional $N=1$ super Yang-Mills in the light cone 
gauge~\cite{Claudson85,DeAlfaro86} and $d$ dimensional lattice linear gauge 
theory~\cite{Matsui88}.

The existence of a local Nicolai map plays a major role in the lattice 
formulation of such models.
Indeed, its discretization provides a recipe for the construction of a
lattice field theory which retains most of the continuum symmetries. In
particular, doubling of the fermions is automatically implemented due to
relations between the boson and fermion 
propagators~\cite{Elitzur,Sakai83}.
Such lattice models can be studied by standard simulation techniques, but
it is interesting to see if the underlying discrete stochastic structure
can be useful for the purpose of numerical computations in a more direct
way.

If one succeeds in solving the stochastic equation $\xi=\xi(\phi)$ then
uncorrelated $\phi$ configurations may be obtained by generating $\xi$
samples. Moreover, explicit fermion fields can be avoided because their
correlation functions may be expressed by means of the so-called stochastic
identities~\cite{DeAlfaro86,Parisi}
in terms of statistical correlations of the solution $\phi(\xi)$
and the random field $\xi$.

The actual implementation of this programme faces three 
difficulties: (i) 
specific boundary conditions must be imposed to preserve supersymmetry
making highly non trivial to determine $\phi(\xi)$ (ii) 
there may be more than one solution (iii) 
the stochastic equation may admit no 
thermal equilibrium~\cite{Cecotti2,Parisi}. 
In other words, when regarded as an evolution 
equation for $\phi$, unbounded
solutions may appear with possible 
instability problems in the numerical simulation.

In this paper, we discuss the above three problems in 
$0+1$ and $1+1$ dimensional specific examples. We study general properties
of the continuum and discretized Nicolai map and perform explicit numerical
simulations to check the feasibility of the method.

%
%

\section{Stochastic Equations and Supersymmetry}

Let $\phi_\alpha(t)$, $t\in [0,\beta]$, be a time dependent field obeying
the equation 
\be
\label{langevin}
\xi_\alpha(t) = \frac{d}{dt} \phi_\alpha(t) +W_\alpha(\phi),
\qquad W_\alpha(\phi) =
\derf{W}{\phi_\alpha},
\ee
where $\xi_\alpha$ is a gaussian white noise
\be
\langle \xi_\alpha(t)\xi_\beta(t')\rangle = \delta_{\alpha\beta}
\delta(t-t'),
\ee
and $W(\phi)$ is an arbitrary potential function. For the moment we do not
specify the boundary conditions in~(\ref{langevin}) which make the problem
well posed. The field index $\alpha$ may include spatial variables which we
assume to vary in a finite volume. Equation~(\ref{langevin}) may be
regarded as a functional change of variable $\phi\to \xi$ which can be
inverted in a certain range of $\xi$ giving rise to many branches
$\phi=\phi^{(n)}(\xi)$, $n=1,\dots, N(\xi)$.

As shown in~\cite{Parisi}, 
the field model with periodic boundary conditions and
classical action
\be
S = \int_0^\beta dt\left(\frac{1}{2}
(\dot\phi_\alpha +W_\alpha)^2+
\psib_\alpha\der{\xi_\alpha}{\phi_\beta}\psi_\beta \right),
\qquad
\phi(0) = \phi(\beta), 
\qquad
\psi(0) = \psi(\beta),
\ee
is $N=2$ supersymmetric and supertraces can be computed as stochastic
averages
\begin{eqnarray}
\label{str}
\mbox{Str}\left[\Omega(\phi)e^{-\beta H}\right] &=& 
\mbox{Tr}\left[\Omega(\phi)e^{-\beta
H}(-1)^F\right] = \\ 
&=& \int \ddd\phi\ddd\psi\ddd\psib\ e^{-S}\Omega(\phi) = 
\sum_n\int\ddd\xi e^{-S(\xi)}\Omega(\phi^{(n)}(\xi)) 
\left. \mbox{sgn}\det\der{\xi}{\phi}\right|_{\phi=\phi^{(n)}(\xi)}, \nonumber
\end{eqnarray}
where $F$ is the fermion number, $H$ is the hamiltonian and $S(\xi) = \frac
1 2 \int_0^\beta \xi^2 dt$.

Equation~(\ref{str}) may be exploited if one is able to solve numerically
\eq{langevin} with periodic boundary conditions $\phi(0) = \phi(\beta)$.

\section{The Nicolai Map in the Continuum}
\subsection{$0+1$ Dimensions, Supersymmetric Quantum Mechanics}
\label{sub:qm}

In the case of SUSY Quantum Mechanics~\cite{Witten1}, 
Equation~(\ref{langevin}) is
simply
\begin{eqnarray}
\label{problem}
\dot q &=& f(q) + \xi, \\
q(0) &=& q(\beta), \nonumber
\end{eqnarray}
and the Hamiltonian $H$ appearing in~(\ref{str}) is 
\be
H = \frac 1 2 p^2 + \frac 1 2 f(q)^2 - \frac 1 2 f'(q) \sigma_3,\qquad
p = -i\frac{d}{dq}, \qquad
\sigma_3 = \left(\begin{array}{cc} 1 & 0 \\ 0 & -1 \end{array}\right) .
\ee
Some properties of~(\ref{problem}) do not depend on $\xi$ being a random
field. For this reason, we begin by regarding~(\ref{problem}) as an inner
map in the space $\smooth$ of periodic smooth functions in $[0,\beta]$.

The Jacobian of Nicolai map can be computed explicitly~\cite{Gozzi} and it is 
\be
\label{Determinant}
\det\left(\dd{}{t} + f^\prime(q)\right) = 
c\cdot \sinh\left(\half \int_0^\beta dt
f^\prime(q(t))\right) .
\ee
The number $N(\xi)$ of solutions $\phi$ for a given $\xi$ is an integer
valued function of $\xi$ with possible jumps across the critical manifold
\be
\label{MC}
{\cal M}_c = \left\{q(t)\in \smooth\ | \ 
\int_0^\beta f'(q(t)) dt = 0\right\}.
\ee
The geometry of ${\cal M}_c$ provides informations on $N(\xi)$; for
instance, suppose that $\smooth/{\cal M}_c$ is simply connected, then
$N(\xi)$ must be constant. To evaluate it, we choose $\xi$ constant and
deduce from 
\be
\dot q = f(q) + \xi\quad \Rightarrow\quad (\dot q)^2 = f(q)\dot q + \xi \dot q
\quad\Rightarrow\quad \int_0^\beta (\dot q)^2 dt = 0,
\ee
that $q$ is also constant and, therefore, given by the roots
of the equation 
\be
f(q) + \xi = 0 .
\ee
A particularly interesting case is that of $f'>0$, which implies 
${\cal M}_c =\emptyset$. In this case, in the open problem
\begin{eqnarray}
\label{open}
\der{}{t} q(t, q_0) &=& f(q(t, q_0)) + \xi(t), \\
q(0, q_0) &=& q_0, \nonumber
\end{eqnarray}
we have 
\be
\der{q(t,q_0)}{q_0} = \exp\int_0^t f'(q(z,q_0)) dz > 1 ,
\ee
so that $\Delta(t, q_0) = q(t, q_0)-q_0$ is a monotone function of $q_0$
and we conclude that if a solution exists then it must be unique.

To see this arguments working in particular cases, let us consider what
happens when $f(q) = \mu q^n$ with $\mu>0$ and $n=1, 2, 3$.

\vskip 0.2cm
\noindent\underline{$n=1$}
\vskip 0.2cm

This is the simplest case. Indeed, $q(\beta, q_0)$ is a 
linear function of $q_0$ and 
the problem 
\baa
\dot q &=& \mu q +\xi, \\
q(0) &=& q(\beta), \nonumber
\eaa
has the unique periodic solution 
\baa
q(t) &=& e^{\mu t} q(0) + e^{\mu t}\int_0^t e^{-\mu z }\xi(z) dz, \\
q(0) &=& \frac{1}{e^{-\mu\beta}-1}\int_0^\beta e^{-\mu t}\xi(t) dt . \nonumber
\eaa

\vskip 0.2cm
\noindent\underline{$n=2$}
\vskip 0.2cm

In this case, $\smooth/{\cal M}_c$ is not simply connected and, moreover,
we cannot solve explicitly the problem
\baa
\label{riccati}
\dot q &=& \mu q^2 +\xi, \\
q(0) &=& q(\beta) . \nonumber
\eaa
Motivated by what happens when $q$ and $\xi$ are constant we guess that, 
in general, there cannot be solutions for all $\xi$ and that, when there are
solutions, 
they come in pairs. This follows also from 
the constraint on $\xi$
\be
\int_0^\beta \xi(t)dt = -\mu \int_0^\beta q^2(t)dt \le 0,
\ee
which excludes some $\xi$ and from the fact that 
\eq{riccati} is a Riccati equation. 
If $q_1(t)$ is one particular solution for a given $\xi(t)$ then
another solution is $q_2(t)$ where
\ba
q_2(t) &=& q_1(t) + \frac{1}{w(t)}, \\
\dot{w} &+& 2 q_1(t) w = -1,\qquad w(0) = w(\beta) .
\ea
In terms of 
\be
F(t) = \exp\left(2\int_0^t q_1(\alpha) d\alpha\right),
\ee
the function $w$ satisfies 
\be
\der{}{t}(F(t) w(t)) = -F(t),
\ee
and the periodicity condition $w(\beta) = w(0)$ gives
\be
w(0) = -\frac{1}{F(\beta)-1}\int_0^\beta F(\alpha)d\alpha .
\ee
Hence, a second solution $q_2$ can always be found except when 
$F(\beta)=1$ namely on the critical manifold
\be
\int_0^\beta q(t)dt = 0,
\ee
of the map.

\vskip 0.2cm
\noindent\underline{$n=3$}
\vskip 0.2cm

This is a case where $f'(q) = 3\mu q^2$ is always positive. The difference 
$\Delta(\beta, q_0)$ is a monotonically increasing function and 
it is easy to prove that 
\be
\lim_{q_0\to \pm\infty} \Delta(\beta, q_0) = \pm\infty,
\ee
(Strictly speaking this holds only at the discrete level where blowing
solutions do not appear). From this remarks it follows that there is a
unique periodic solution for each periodic $\xi(t)$.

In the above analysis, $\xi$ was assumed to be smooth. When $\xi$ is a
white noise it gives fluctuations $\delta q$ around the $\xi=0$ solutions,
namely $q=q^*$ with $f(q^*)=0$. The size of $\delta q$ depends on
$f'(q^*)$ and $\beta$. Due to periodicity, we expect it to 
approach a constant when
$\beta\to \infty$ and diverge as $\beta\to 0$ with no regards to the
stability of the fixed point $q=q^*$.

An illustrative solvable example is that of $f(q)=\mu q$. From
\be
q(0) = \frac{1}{e^{-\mu\beta}-1}\int_0^\beta e^{-\mu t}\xi(t) dt ,
\ee
we obtain 
\be
\langle q(0)^2\rangle = \frac{1}{2\mu} \frac{1-e^{-2\mu\beta}}
{(1-e^{-\mu\beta})^2},
\ee
and indeed we find 
\be
\langle q(0)^2\rangle \stackrel{\beta\to 0}{\sim} \frac{1}{\mu^2\beta},
\qquad 
\langle q(0)^2\rangle \stackrel{\beta\to +\infty}
{\longrightarrow} \frac{1}{2|\mu|}.
\ee
Here $q^*=0$ and the parameter $\mu$ is just 
$f'(q^*)$ namely what we can call the tree level mass. In a
numerical simulation, in the $\beta\to\infty$ limit, we expect to have one
solution for each fixed point with large deviations depressed 
exponentially by potential barriers controlled by 
parameters like $\mu$. In these regimes (assuming $f'(q^*)\neq 0$)
we can compute the sign of the Jacobian
$\det\partial \xi/\partial q$ at each $q=q^*$ and use it in a neighbourhood of
that point.

\subsection{$1+1$ Dimensions, WZ Models}
\label{sub:wz}

In $1+1$ dimensions, let $z=x_1 + i x_2$ and $\phi(z, \zb)$ be a complex
field. Equation~(\ref{langevin}) takes the form 
\be
\label{langevincomplex}
2 \der{\phi}{z} = \overline{f(\phi)} + \eta ,
\ee
where $\eta = \eta_1 + i \eta_2$ with $\eta_1$ and $\eta_2$ 
real independent white noises and $f(\phi) = u(\phi)+i v(\phi)$ an 
arbitrary holomorphic function of $\phi$. 
The reason for the peculiar structure of~(\ref{langevincomplex}) is 
that it guarantees that the 
associated field model (the so-called WZ model) is Lorentz
covariant~\cite{Parisi}. The explicit form
of~(\ref{langevincomplex}) is 
\bt{complexexplicit}
\partial_1\phi_1 +\partial_2\phi_2 &=& u(\phi_1, \phi_2) + \eta_1 ,\\
\partial_1\phi_2 - \partial_2\phi_1 &=& -v(\phi_1, \phi_2) + \eta_2 .
\et
The numerical integration of~(\ref{complexexplicit}) is difficult 
because even in the simplest cases the associated random flow 
does not admit an 
equilibrium distribution. To see this, it is enough to consider 
a homogeneous (independent on $x_2$) solution of~(\ref{complexexplicit})
without noise. It must satisfy
\be
\dd{\phi}{x_1} = \overline{f(\phi)}.
\ee
If $F(\phi)$ is a primitive of $f(\phi)$ then the quantity
$H = \mbox{Im} F(\phi)$ is a constant of motion since
\be
\dd{H}{x_1} = \mbox{Im}\left( f(\phi)\dd{\phi}{x_1}\right) = \mbox{Im}|f|^2
= 0.
\ee
If we consider a function $f(z)$ with asymptotic power behaviour $\sim z^n$
we see that the level curves of $H$ are not closed and 
equilibrium cannot be reached.

As in the $0+1$ dimensional case, for a constant $\eta$ 
we have (by periodicity)
\be
2 \der{\phi}{z} = \overline{f(\phi)} + \eta\ \Rightarrow\ 
2 \left|\der{\phi}{z}\right|^2 = \overline{f(\phi)\der{\phi}{z}} + 
\eta \overline{\der{\phi}{z}} \ \Rightarrow\ 
\int_0^\beta dx_1\int_0^\sigma dx_2 \left|\der{\phi}{z}\right|^2 =0 ,
\ee
which implies that $\phi$ is a constant obtained from the equation  
\be
f(\phi) + \overline{\eta} = 0 .
\ee
In particular, when $\eta\to 0$, $\phi$ tends to one of
the zeroes of $f(\phi)$. As in the $0+1$ dimensional case, the Nicolai map
is not singular at $\phi=\phi^*$ with $f(\phi^*)=0$ and $f'(\phi^*)\neq 0$.
Actually, an infinitesimal periodic zero mode $\lambda$ such that 
$\phi = \phi^* + \lambda$ is a solution of~(\ref{langevincomplex})
must satisfy
\be
\label{zeromode}
2\der{\lambda}{z} = \overline{f'(\phi^*) \lambda},
\ee
and therefore
\be
(4\partial\bar\partial-|f'(\phi^*)|^2) \lambda = 0 , 
\qquad \Rightarrow \qquad
\int dx_1 dx_2
\left( |\nabla \lambda|^2 + |f'(\phi^*)|^2 |\lambda|^2\right) = 0,
\ee
giving $\lambda\equiv 0$.

\section{The Discrete Nicolai map}

When time is discretized the interval $[0,\beta]$ is divided into $N$ sub
intervals and the Nicolai map is a change of variables 
\be
\label{change}
(\xi_0, \dots, \xi_{N-1})\leftrightarrow (q_0, \dots, q_{N-1}).
\ee
where $q_n$ and $\xi_n$ are the values of $q$ and $\xi$ at times 
$t_n = n\beta/N$. The next step is
the choice of a definite discretization scheme of~(\ref{langevin}). Then,
one has (i) a discrete stochastic equation whose numerical 
properties must be studied and (ii) a precise fermion action to be analyzed
from the point of view of the fermion doubling problem.
In this Section we address these problems as well as the use of the 
so-called stochastic identities to compute fermion correlation functions
without introducing Grassmann fields at all.

\subsection{Choice of the Discretization Procedure}

For simplicity, let us consider a single component field $q(t)$ in $0+1$
dimensions. The Langevin equation 
\baa
\dot q &=& f(q) + \xi, \\
\langle \xi(t)\xi(t')\rangle &=& D(q(t))\delta(t-t'), \nonumber
\eaa
where we allowed a $q$ dependent covariance, may be discretized according
to 
\be
\label{langevindiscrete}
q_{n+1} = q_n + \epsilon\left[
\alpha f(q_n)  + (1-\alpha) f(q_{n+1})
\right] + \sqrt{\epsilon} D^{1/2}(\alpha q_n + (1-\alpha) q_{n+1})\xi_n,
\ee
where the constant parameter $\alpha$ takes into account the
ambiguity in the evalutation of $f$ and $D$. We recall that $\alpha=1$ 
and $\alpha=1/2$ are usually referred to as the Ito and Stratonovich
discretization schemes~\cite{Ezawa85}. 
For non constant $D(q)$, the value of $\alpha$ is
important. Actually, in the $\epsilon\to 0$ limit we can 
replace~(\ref{langevindiscrete}) by 
\be
q_{n+1} = q_n + \epsilon \left[
f(q_n)  + (1-\alpha) D'(q_n) \xi_n^2
\right] + \sqrt{\epsilon} D^{1/2}(q_n)\ \xi_n + \cdots , 
\ee
and read the associated Fokker-Planck equation 
\be
\der{P}{t} = \frac 1 2 \der{}{q}\left(D^{1-\alpha}\der{}{q}\left
(D^\alpha P\right)\right)-\der{}{q}(f P) .
\ee 
In our case, however, $D\equiv 1$ is a constant and the
same continuum limit is obtained for each value of $\alpha$.
However, the equivalence of the two 
discretizations is not obvious at first sight
because the Jacobian of the change of variables~(\ref{change})
depends on $\alpha$. Actually, the Jacobian of the transformation 
$\{q\}\to \{\xi\}$ expressed by~(\ref{langevindiscrete})
with periodic boundary conditions $\xi_N=\xi_0$, $q_N=q_0$
is proportional to 
\be
J = \det\left(
\begin{array}{ccccc}
-1-\alpha\epsilon f_0^\prime & 1-(1-\alpha)\epsilon f_1^\prime & 0 & \cdots
& 0 \\
0 & -1-\alpha\epsilon f_1^\prime & 1-(1-\alpha)\epsilon f_2^\prime & \cdots
& 0 \\
\cdots & & & & \\
1-(1-\alpha)\epsilon f_0^\prime & 0 & 0 &  \cdots & 
-1-\alpha\epsilon f_{N-1}^\prime 
\end{array}
\right), \qquad f_n \equiv f(q_n),
\ee
which we can compute 
\baa
J &=& \prod_{k=0}^{N-1} \left(-1-\alpha\epsilon f_k^\prime\right) +
(-1)^{N-1} \prod_{k=0}^{N-1} \left(1-(1-\alpha) \epsilon f_k^\prime\right) \\
&\stackrel{\epsilon\to 0}{\sim}& 
e^{\alpha\int f'(q(t))dt}\left(1-e^{-\int f'(q(t)) dt}\right). \nonumber
\eaa
On the other hand, in the discrete gaussian action 
\baa
\frac 1 2 \sum_n \xi_n^2 &=& \\ \nonumber
&=& \frac{1}{2\epsilon}\sum\left\{q_{n+1}-q_n-\epsilon\left[\alpha f_n +
(1-\alpha) f_{n+1}\right]\right\} = \\
&=& \frac{1}{2\epsilon}\sum_n\left\{(q_{n+1}-q_n)^2+\epsilon^2
f_n^2\right\}-\sum_n (q_{n+1}-q_n)(\alpha f_n + (1-\alpha) f_{n+1}) .
\nonumber
\eaa
We need some care to take the limit $\epsilon\to 0$ in the last term.
However, if we write 
\baa
\lefteqn{\sum_n (q_{n+1}-q_n)(\alpha f_n + (1-\alpha) f_{n+1}) = } && \\
&& \sum_n (q_{n+1}-q_n)\left(\frac 1 2 (f_n+f_{n+1}) + \frac{1-2\alpha}{2}
(f_{n+1}-f_n)\right), \nonumber
\eaa
then the correct continuum limit can be read: it depends on $\alpha$ and is
\be
\int f(q)dq + \left(\frac 1 2 -\alpha \right) \int f'(q(t)) dt .
\ee
Putting this result together with $J$ we find a final expression 
which is independent from $\alpha$ and reproduces precisely~\eq{Determinant}.

In numerical simulations, the most convenient choice of $\alpha$ is 
$\alpha=1$ because in this case $q_{n+1}$ may be evaluated directly from
$q_n$ without solving any equation.

\subsection{Doubling Problem}

As is well known, naive discrete fermion actions are plagued by the 
doubling problem~\cite{W77}. 
In a formulation based on a discrete Nicolai map 
the fermion action is constrained by the bosonic one (another manifestation
of supersymmetry) and should be free from doublers. In this section 
we briefly explain how the discrete Nicolai map accomplishes this task by 
implementing naturally
Wilson type lattice fermions. To see this, 
let us adopt the standard notation 
\baa
\nabla^+ f_n &=& f_{n+1}-f_n, \qquad
\nabla^- f_n = f_n-f_{n-1}, \nonumber \\
\nabla^S f_n &=& \frac 1 2 (\nabla^+ + \nabla^-) f_n, \qquad 
\nabla^A f_n = \frac 1 2 (\nabla^+ - \nabla^-) f_n. \nonumber
\eaa
The doubling problem is related to the kinetic term only and 
in this Section we set $f\equiv 0$. We begin with the $0+1$ dimensional
case: the Ito discretization of equation $\dot q = \xi$,
is based on the replacement $\dot q \to \nabla^+ q$.
This fixes the fermion matrix $\partial \xi/\partial q$ and leads
to the fermion action
\be
\label{fermionfirst}
S_F = \sum_n \psib_n (\psi_{n+1} - \psi_n).
\ee
On the other hand, since $\nabla^+ = \nabla^S + \nabla^A$, we can identify 
in~\eq{fermionfirst} the naive fermion action plus a Wilson term
\be
S_F = \frac 1 2 \sum_n \psib_n (\psi_{n+1} - \psi_{n-1}) +
\frac 1 2 \sum_n \psib_n (\psi_{n+1}-2\psi_n + \psi_{n-1}).
\ee
A similar mechanisms happens in $1+1$ dimensions. For instance, 
the discrete Nicolai map proposed in~\cite{Sakai83} for the WZ models
is a cubic symmetric version of $\dot q \to \nabla^+ q$.
The continuum map
\be
\partial_1\phi + i\sigma_2\partial_2\phi = \xi,
\qquad \phi =
\left(\begin{array}{c}\phi_1 \\ \phi_2 \end{array}\right),
\qquad \xi =
\left(\begin{array}{c}\xi_1 \\ \xi_2 \end{array}\right), \qquad
\sigma_2 = \left(\begin{array}{cc} 0 & -i \\ i & 0 \end{array}\right),
\ee
may be discretized as~\footnote{
With this choice of $\nabla^+$ and $\nabla^-$ the cross terms in $\xi_1^2 +
\xi_2^2$ cancel.
}
\ba
(\nabla_1^+ + \nabla_2^A) \phi_1 + \nabla_2^S\phi_2 &=& \xi_1, \\
(\nabla_1^- - \nabla_2^A) \phi_2 - \nabla_2^S\phi_1 &=& \xi_2, 
\ea
which, using $\nabla^\pm = \nabla^S\pm \nabla^A$, 
leads to the fermion action 
\be
S_F = \sum_n \psib_n 
\left[\nabla_1^S+ i\sigma_2\nabla_2^S + 
\sigma_3(\nabla_1^A+\nabla_2^A)\right]
\psi ,
\ee
namely (after a redefinition of $\psi$) 
the naive discrete action plus a Wilson term.

\subsection{Some Exact Properties of the Discrete Maps}

In~\refsec{sub:qm} and~\refsec{sub:wz} we discussed some analytical tools
for the study of the existence of periodic solutions of the Langevin
equation as well as for the determination of their number $N(\xi)$.

Some of those conclusions hold also in the discrete case.
Let us begin with the $0+1$ dimensional case. The Ito discretization 
\be
\label{dlangevin}
q_{n+1} = q_n + \epsilon f(q_n) + \sqrt\epsilon \xi_n,
\ee
of the open problem~(\ref{open}) gives
\be
\frac{dq_N}{dq_0} = \prod_n (1+\epsilon f'(q_n)) .
\ee
Hence, in the case $f'(q)>0$ we have
\be
\frac{d}{dq_0} (q_N-q_0) > 0,
\ee
and (at least for asymptotic $f\sim q^{2n+1}$ with positive $n$) it is easy
to prove the existence of a unique periodic sequence $\{q_n\}$ for each
periodic $\{\xi_n\}$.


Concerning the WZ models, we now show that also in the discrete
equations there are not zero modes
when $\phi$ is constant. The equation for the zero mode $\lambda$ 
is~(see \eq{zeromode})
\be
\label{zeromodedisc}
\left[
\nabla_1^S + i\sigma_2\nabla_2^S+\sigma_3(\nabla_1^A +
\nabla_2^A-u_1)-\sigma_1 u_2
\right]\lambda=0,
\qquad u_i =
\left. \der{u}{\phi_i}\right|_{\phi=\phi_0},
\ee
Expanding the periodic $\lambda$ in Fourier series 
\be
\lambda_i(x_1, x_2) = 
\sum_{k_1, k_2} c_{i, k_1, k_2} e^{i(k_1 x_1 + k_2 x_2)} ,
\ee
we find the determinant of the operator in~(\ref{zeromodedisc})
\baa
\lefteqn{\det\left[
\nabla_1^S + i\sigma_2\nabla_2^S+\sigma_3(\nabla_1^A +
\nabla_2^A-u_1)-\sigma_1 u_2
\right] =} && \\
&=&  \prod_{k_1, k_2} \left[-\sin^2 k_1-\sin^2 k_2-u_2^2 -\left(u_1 +
2\sin^2\frac{k_1}{2}+ 2\sin^2\frac{k_2}{2}\right)^2 \right], \nonumber
\eaa
which is not zero unless $u_1=u_2=0$.

\subsection{Stochastic Identities}

The stochastic identities relate fermion correlation functions to
stochastic averages involving the solution $q$ of the Langevin equation and
the noise $\xi$. In this paper we shall need only
the simplest of them. To prove it at the discrete level we write 
\baa
\vev{\Omega(q)\xi_\alpha} &=& \nonumber
\int \frac{d\xi_0}{\sqrt{2\pi}}\cdots\frac{d\xi_{N-1}}{\sqrt{2\pi}}
e^{-\half(\xi_0^2+\cdots+\xi_{N-1}^2)} \Omega(q(\xi)) \xi_\alpha = \\
&=& -\int \frac{d\xi_0}{\sqrt{2\pi}} \cdots \frac{
\left(de^{-\half\xi_\alpha^2}\right)}{\sqrt{2\pi}} 
\cdots \frac{d\xi_{N-1}}{\sqrt{2\pi}} \ \Omega(q(\xi)) = \\
&=& \int \frac{d\xi_0}{\sqrt{2\pi}}\cdots\frac{d\xi_{N-1}}{\sqrt{2\pi}}
e^{-\half(\xi_0^2+\cdots+\xi_{N-1}^2)} \der{}{\xi_\alpha} \Omega(q(\xi)) = 
\nonumber \\
&=& \vev{\der{\Omega}{q_\beta}\der{q_\beta}{\xi_\alpha}}
\nonumber
\eaa
However, in terms of the fermion matrix
\be
J_{\alpha\beta} = \der{\xi_\alpha}{q_\beta},
\ee
this means
\be
\vev{\Omega(q) \xi_\alpha} = \vev{\partial_\lambda \Omega(q) 
(J^{-1})_{\lambda\alpha}},
\ee
and, in particular,
\be
\label{stochid}
\vev{q_\alpha\xi_\beta} = \vev{(J^{-1})_{\alpha\beta}} = 
\vev{\psib_\alpha\psi_\beta},
\ee
which expresses the fermion propagator in terms of the $q$-$\xi$ correlation.

\section{Numerical Simulation}

From the very existence of a local Nicolai map and previous discussions
it follows an algorithm for
the numerical computation of supertraces. The first step is the extraction
of the gaussian random numbers $\{\xi_n\}$. Then, let $\phi_n$ be the field
obeying a discretized version of~(\ref{langevin}); we must find the initial
condition $\phi_0$ such that
\be
\label{target}
\Delta(\phi_0) = \Vert \phi_N(\phi_0) - \phi_0 \Vert = 0 .
\ee
In $0+1$ dimensions it is easy to identify all solutions
of~(\ref{target}) as well as the sign of the Jacobian determinant of the
Nicolai map. For the WZ models in $1+1$ the problem is
harder. However, as we have seen, 
at least for large separation of the zeroes $\phi^*$ of
$f(\phi)$ we can expect to have one solution for each $\phi^*$. Therefore
we can use $\{\phi^*\}$ as starting guesses and take for
$\det\partial\xi/\partial q$ its value at $\phi^*$. In the
simplest case $f(\phi) = \mu\phi+g\phi^2$ we have $\phi^*=0,-\mu/g$ and the
above regime is obtained for large $\mu/g$.

If an operator ${\cal O}(\phi)$ is averaged as in~(\ref{str}) over the
realizations of $\{\xi\}$ we obtain an estimate of the supertrace
$\mbox{Str}\left[{\cal O}\exp(-\beta H)\right]$ 
and when $\beta\to +\infty$ we obtain
$\langle 0 | {\cal O} | 0\rangle$ in the case of unbroken SUSY. In the free
$0+1$ dimensional case $f(q) = \mu q$ (and 
similarly in the free WZ models) we can solve analitically the discrete
equations and determine the correlation functions with the
stochastic algorithm or in field theory.

The discrete Langevin equation is 
\be
\label{farfalla}
q_{n+1} = \omega  q_n + \sqrt{\epsilon}\ \xi_n , \qquad
\omega = 1+\epsilon\mu .
\ee
The linear system
\be
\left(
\begin{array}{ccccc}
\omega & -1 & 0 & \cdots & 0 \\
0 & \omega & -1 & \cdots & 0 \\ & & \cdots \\ -1 & 0 & 0 & \cdots &
\omega
\end{array}
\right) 
\left(
\begin{array}{c}
q_0 \\ q_1 \\ \cdots \\ q_{N-1}
\end{array}
\right) = -\sqrt{\epsilon}\ 
\left(
\begin{array}{c}
\xi_0 \\ \xi_1 \\ \cdots \\ \xi_{N-1}
\end{array}
\right)
\ee
is easily solved. The inverse of the first factor is
\be
\frac{1}{\omega^N-1}\cdot
\left(
\begin{array}{cccc}
\omega^{N-1} & \cdots & \omega & 1\\
1 & \cdots & \omega^2 & \omega \\ & \cdots \\
\omega^{N-2} & \cdots & 1 & \omega^{N-1}
\end{array}
\right)
\ee
and therefore the initial value is
\be
q_0 = -\sqrt{\epsilon} \frac{1}{\omega^N-1} \left(\omega^{N-1} \xi_0 +
\cdots + 1\cdot \xi_{N-1}\right) .
\ee
The two-point function is
\be
C_k = \vev{q_0 q_k} = \frac{\epsilon}{(\omega^{N} -
1)^2}\sum_{i=0}^{N-1}
\omega^{i\ {\rm mod}\ N + (i+k)\ {\rm mod}\ N} ,
\ee
and after a straightforward algebra
\be
C_k = \frac{2\epsilon\ \
\omega^{N/2}}{(\omega^2-1)(\omega^N-1)}
\cosh\left(\left(k-\frac{N}{2}\right)\log \omega\right) .
\ee
If we take the limit $N\to\infty$ with $\epsilon = \beta/N$ and introduce
the time variable $\tau = \epsilon k$ we obtain
\be
\label{explicit}
\mbox{Str}(q e^{-\tau H} q e^{-\beta H}) = \lim_{N\to\infty}\vev{q_0 q_k} = 
\frac{1}{2\mu}\frac{\cosh\left(\mu(\tau-\beta/2)\right)}{\sinh(\mu\beta/2)}
\ee
and
\be
\lim_{\beta\to+\infty} \mbox{Str}(q e^{-\tau H} q e^{-\beta H}) = 
\langle 0 | q(0) q(\tau) | 0 \rangle = \frac{1}{2\mu}\exp(-\mu\tau) .
\ee
From the field theoretical point of view, we are computing 
$\mbox{Str}(q e^{-\tau H} q e^{-\beta H})$
with the free action
\be
S = \frac{1}{2}\int_0^\beta dt(\dot{q}^2 + \mu^2 q^2),
\ee
and periodic boundary conditions. The generating functional is
\be
Z = \vev{\exp\left(\int\ d\tau J q\right)} = \exp\left(\int d\tau
d\tau^\prime J(\tau)G(\tau-\tau^\prime) J(\tau^\prime)\right),
\ee
where
\be
G(\tau) = \frac{1}{2\beta}\sum_{k=-\infty}^\infty \frac{1}{\displaystyle
\mu^2 + \frac{4\pi^2k^2}{\beta^2}}\exp\left(-\frac{2\pi i k}{\beta}
\tau\right),
\ee
and
\be
\mbox{Str}(q e^{-\tau H} q e^{-\beta H}) = 2 G(\tau) .
\ee
If we use the summation formula
\be
\sum_{k=-\infty}^\infty\frac{\exp(-i\alpha k)}{k^2+\lambda^2} = 
\frac{\pi}{\lambda\sinh(\pi\lambda)}
\cosh\left(\lambda(\alpha-\pi)\right), \qquad 0<\alpha<2\pi ,
\ee
we obtain again the result of Eq.(\ref{explicit}). A similar computation
can be carried out for the fermionic propagator. From the solution
of~(\ref{farfalla}) we have
\be
\langle q_k\xi_0\rangle \frac 1 {\sqrt{\epsilon}} =
\frac{\omega^{k-1}}{1-\omega^N} ,\qquad
\langle q_0\xi_k\rangle \frac 1 {\sqrt{\epsilon}} =
\frac{\omega^{N-k-1}}{1-\omega^N} ,
\ee
which can be combined to give
\be
\langle q_k \xi_l\rangle\frac{1}{\sqrt{\epsilon}} =
\omega^{k-l-1}\left[\frac 1 {1-\omega^N}-\theta(l-k)\right],
\ee
which has the correct continuum limit
\baa
\langle q(\tau)\xi(\tau')\rangle &=& e^{\mu(\tau-\tau')}\left[
\frac 1 {1-e^{\mu\beta}} -\theta(\tau'-\tau)
\right] = \\
&=& \frac 1 \beta \sum_{n=-\infty}^\infty \frac{1}{\displaystyle
\frac{2\pi i n}{\beta}-\mu} 
\exp\left( \frac{2\pi i n}{\beta}(\tau-\tau') \right),
\nonumber 
\eaa
associated to the fermion propagator in $0+1$ dimensions 
and confirming the stochastic identity.

In the interacting case, $\phi_0$ must be determined by some iterative
algorithm. In $1+1$ dimensions it is fundamental to start from a good
guess. We use the Newton-Raphson algorithm~\cite{Newton} to solve
iteratively the set of non linear equations
\be
\Delta_i = \phi_i^{(N)}-\phi_i^{(0)} = 0,
\ee
where $\phi_i^{(k)}$ is the $i$-th spatial component of $\phi$ at the
$k$-th time slice. The correction $\delta\phi_i^{(0)}$ in 
\be
\phi^{(0)}_i\to\phi^{(0)}_i + \delta\phi^{(0)}_i ,
\ee
is given by 
\be
\der{\Delta_i}{\phi^{(0)}_j}\ \delta\phi^{(0)}_j = -\Delta_i .
\ee
The scheme is made more robust by introducing a relaxation parameter
$\omega_R$ in the update of $\phi_i^{(0)}$
\be
\phi^{(0)}_i\to\phi^{(0)}_i + \omega_{\rm R}\ \delta\phi^{(0)}_i,
 \qquad 0<\omega_{\rm R}<1 .
\ee
Another help against numerical instabilities is to require
\be
\frac{|\delta\phi^{(0)}|}{|\phi^{(N)}|} < \rho \ \ \mbox{or simply}\ \
|\delta\phi^{(0)}| < \rho, 
\ee
where $\rho$ is a minimum correction threshold.  The choice of the 
optimal $\omega_{\rm R}$ and $\rho$  
must be done empirically, but we did not find it to be critical.

Finally, another general trick which is useful to improve numerical
stability is to follow a bootstrap procedure and solve the
problem on a $L\times (T-1)$ lattice to provide a guess for the $L\times T$
problem.

As a numerical test of the algorithm we measure the boson and
fermion propagators in simple $0+1$ and $1+1$ dimensional cases. In $1+1$,
to gain statistics, we average over the spatial dimension and sum over all
pairs of time slices with fixed temporal separation. Moreover, we
explicitly symmetrize the propagators under $\tau\to \beta-\tau$. 

The simplest interacting system in $0+1$ is 
\be
f(q) = -\mu q -g q^3,\qquad \mu, g > 0 .
\ee
where dynamical breaking of supersymmetry does not occur. 
In Tab.~(\ref{tavolaQM}) we show the lightest mass as a function of $g$ at 
$\mu=4$, $\beta=5$ on a $T=200$ lattice. 
It is obtained by fitting the boson and fermion propagators where the
latter is computed by means of the stochastic identity~(\ref{stochid}). We
also show the ${\cal O}(g^2)$ perturbative value
\be
E_1 = \mu +\frac{3}{2\mu} g - \frac{9}{2\mu^3} g^2 + O(g^3) 
\ee
In~\reffig{fig:1}, 
just to give an example at the critical point $\mu=0$ we plot
the two propagators at $g=1$ and $\mu=0$ computed with 
$\beta=10$ on a $T = 100$ lattice.  As expected,
the slope of the logarithmic plots is the same.

In $1+1$ dimensions, we simulate the WZ model with 
$f(\phi) = \mu\phi + g\phi^2$. 
In~\reffig{fig:2}, we show the boson and fermion propagators evaluated at 
$\mu=4$, $g=0.1$ on a $20\times 90$ lattice with 
$\epsilon_s = 0.1$ and $\epsilon_t = 0.01$ where $\epsilon_s$ and
$\epsilon_t$ are the space and time discretization steps. The 
continuous line is the fit with the Ansatz
\be
C(\tau) = A_0 \cosh\left[\mu\left(\tau - \half\beta\right)\right] .
\ee
On a $20\times 50$ lattice we have varied $g$ with the results reported 
in Tab.~(\ref{tavola}) together with the one loop value of the doublet mass
which is
\be
m(g) = \mu -\frac{2}{3\sqrt{3}}\frac{g^2}{\mu} .
\ee
Finite time step errors can be investigated in a first approximation 
by studying numerically
the finite lattice propagator integrated over space. This, in standard 
notation, is
\be
D(k) = \sum_{n_t} \exp(k \ \epsilon_t p_t) \frac{1}{\hat{p}_t^2 + \mu^2}
\ee
where
\be
p_t = \frac{2\pi}{\beta} n_t \qquad n_t = 0 \cdots T-1\qquad {\rm
and}\qquad
\hat{p_t} = \frac{2}{\epsilon_t} \sin\half \epsilon_t p_t .
\ee
One can study at fixed $\beta$, the difference in the propagator as
$\epsilon_t$ is varied.  We checked that at $\mu=4$, $\epsilon_t = 0.01$
and $T=50$ the finite step effects are negligible.

We also remark that with this particular choice of $f(\phi)$ we can sample
a single zero ($\phi^* = 0$) without violating supersymmetry. The reason is
that the shifted field $\tilde\phi = (\phi_1-\mu/g, \phi_2)$ obeys the same
equations as $\phi$ but with $\mu\to -\mu$. For $\langle \phi\phi\rangle$
and the symmetrized $\langle \psib\psi\rangle$ 
this change has no consequences.

\section{Remarks and conclusions}
\label{sec:conclusions}

In this paper we have shown that the existence of a local Nicolai map in
supersymmetric models has useful consequences for numerical computations. It
allows the formulation of a simulation algorithm which generates
statistically independent field configurations by solving a Langevin
equation with periodic boundary conditions. The so called stochastic
identities can be exploited to avoid Grassmann fields. The method is
feasible and consistent numerical
results are obtained in $0+1$ dimensions and also in $1+1$ WZ models
even if with some constraint in parameter space. Further developments are
possible in the direction of more robust integration schemes for the
Langevin equation as well as in the application to more realistic
models. In particular, work is in progress on cases where the Nicolai map
is determined perturbatively~\cite{FLMR85} and on
$QCD_4$~\cite{Claudson85,DeAlfaro86} where the Jacobian of the local 
Nicolai map is constant.

\figlab{fig:1}{Boson and fermion propagators for the $0+1$
dimensional model with drift $f(q) = -q^3$ at $\beta=10.0$
on a $T = 100$ lattice. Apart from the different normalization
the slopes of the two logarithmic plots are equal.}

\figlab{fig:2}{Boson and Fermion propagators for the 
WZ model with drift $f(\phi) = \mu\phi + g\phi^2$ at the point $\mu=4$, 
$g=0.1$ on a $20\times 90$ lattice with space and time steps 
$\epsilon_s = 0.1$, $\epsilon_t = 0.01$.}

\begin{table}
\caption{Lightest boson ($m_B$) and fermion ($m_F$) masses as functions of 
$g$ at $\mu=4$, 
$\beta=5$ on a $T=200$ lattice in the $0+1$ dimensional model with
drift $f(q) = -\mu q -g q^3$}
\label{tavolaQM}
\begin{center}
\begin{tabular}{cccc}
$g$ & $m_B$ & $m_F$ & $m_{pert}$\\
\tableline
0.0 & 4.00(3) &  4.01(3)  & 4.0000 \\
0.1 & 4.04(3) &  4.04(3)  & 4.0368 \\
0.2 & 4.07(3) &  4.08(3)  & 4.0750 \\
0.3 & 4.11(3) &  4.11(3)  & 4.1062 \\
0.4 & 4.14(3) &  4.14(3)  & 4.1388 \\
0.5 & 4.17(3) &  4.18(3)  & 4.1699 \\
0.6 & 4.20(3) &  4.21(3)  & 4.1997 \\
0.7 & 4.23(3) &  4.24(3)  & 4.2281 \\
0.8 & 4.26(3) &  4.27(3)  & 4.2550 \\
\end{tabular}
\end{center}
\end{table}

\begin{table}
\caption{
Lightest boson ($m_B$) and fermion ($m_F$) masses as functions of 
$g$ at $\mu=4$, $\beta=0.5$ on a $20\times 50$ lattice for the 
WZ model associated to $f(\phi) = \mu\phi + g\phi^2$.}
\label{tavola}
\begin{center}
\begin{tabular}{cccc}
$g$ & $m_B$ & $m_F$ & $m_{pert}$\\
\tableline
0.0 & 4.00(5)  & 4.00(5)  & 4.000     \\
0.1 & 4.00(5)  & 3.99(5)  & 3.999     \\
0.4 & 3.99(5)  & 3.99(5)  & 3.985     \\
0.6 & 3.98(5)  & 3.98(5)  & 3.965     \\
1.2 & 3.96(5)  & 3.97(5)  & 3.859     \\
\end{tabular}
\end{center}
\end{table}

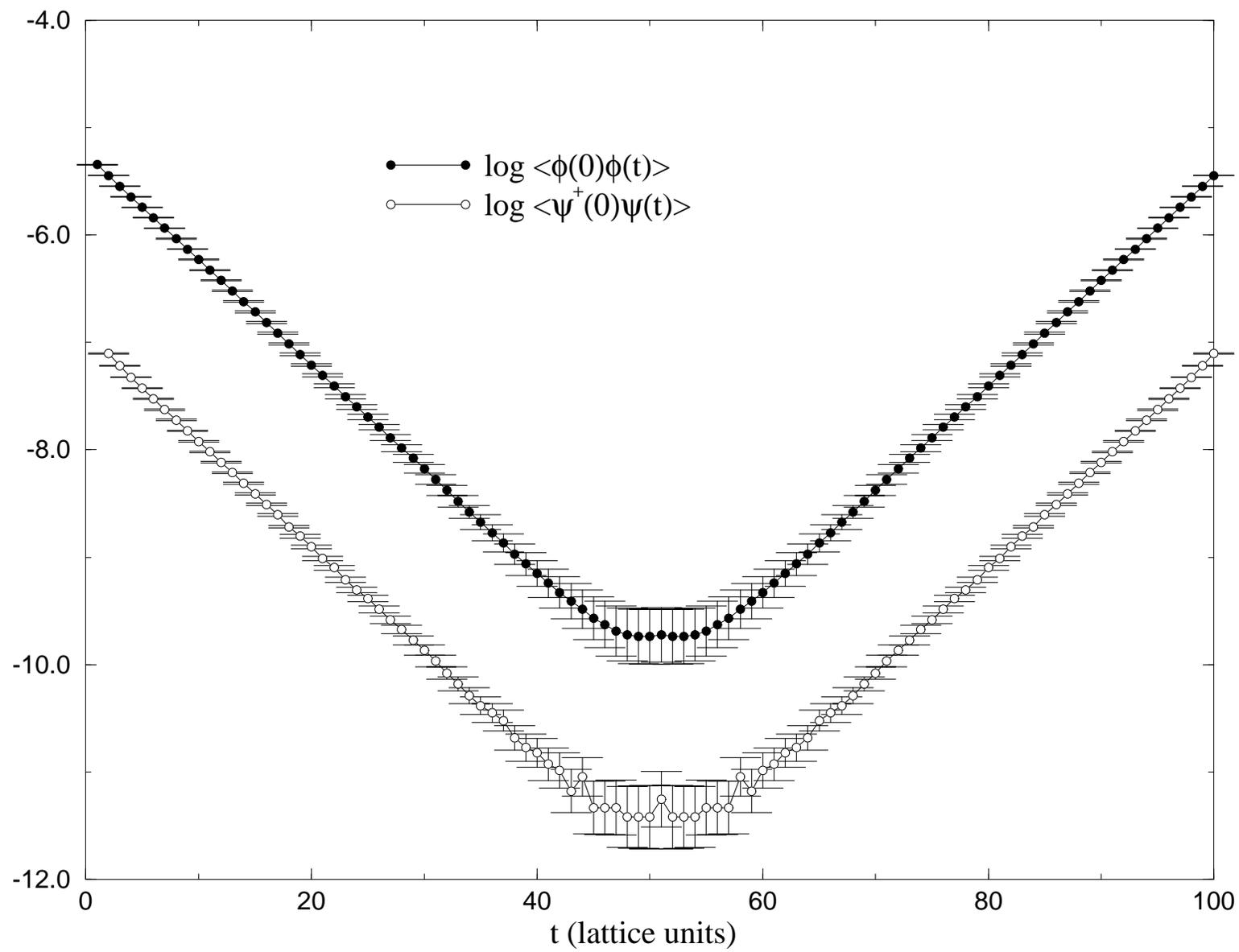
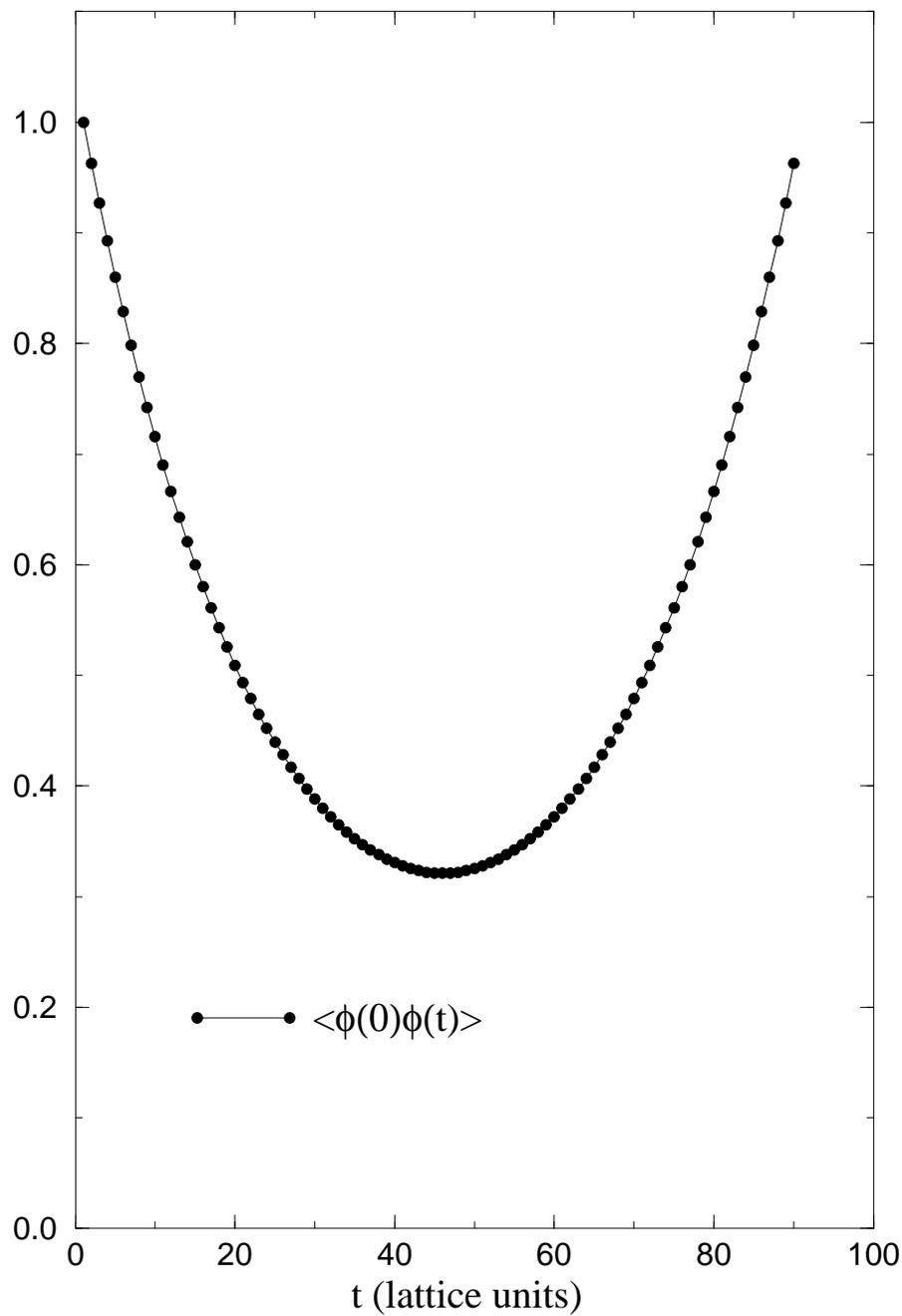
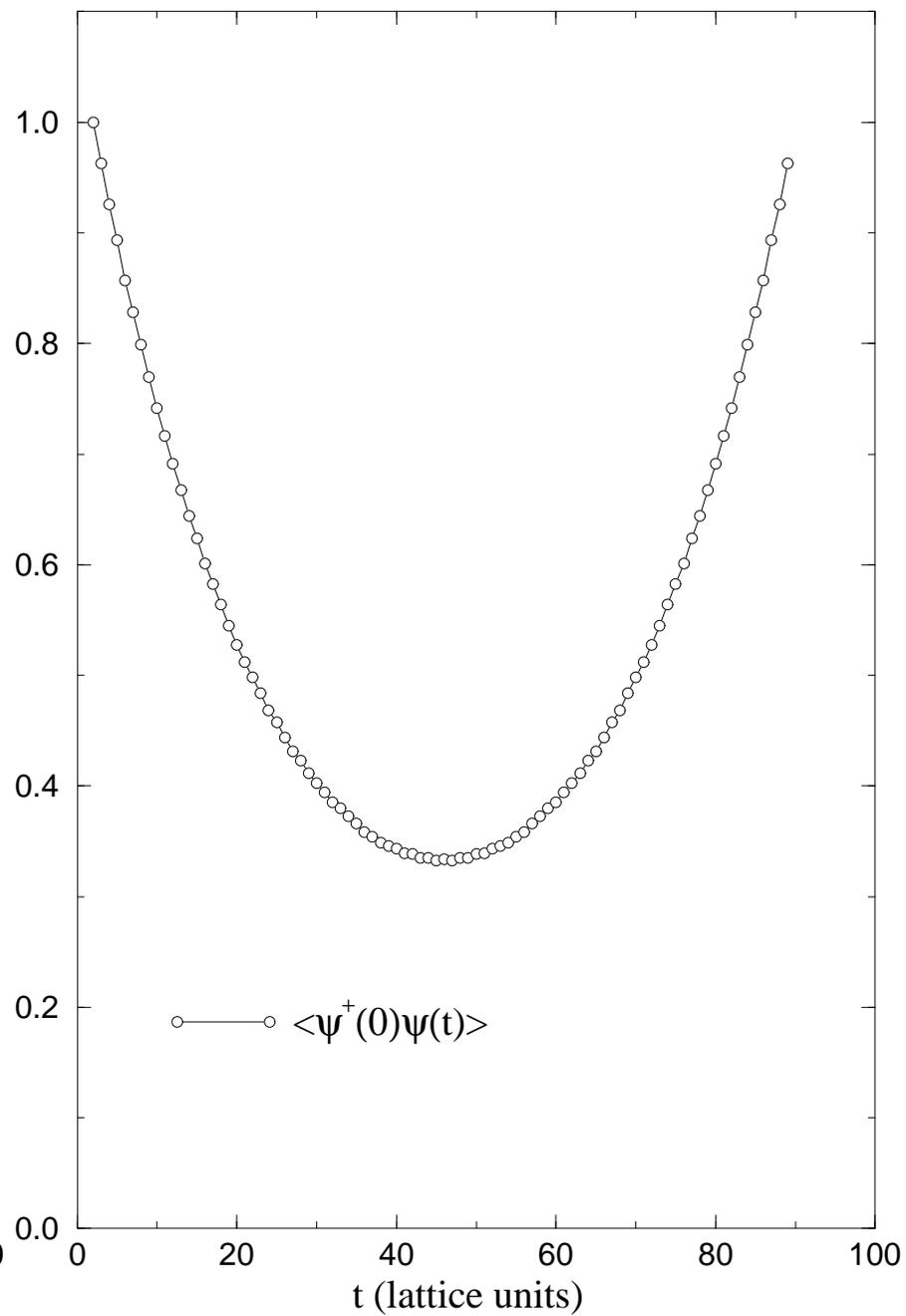
\end{document}